\documentclass[pra,12pt,tightenlines,nofootinbib,floatfix]{revtex4} 
\pdfoutput=1
\usepackage{amsmath}
\usepackage{graphics, graphicx}
\usepackage{epsfig}
\usepackage{ulem}
\usepackage{color}
\usepackage{datetime} 

\begin{document} 




\begin{abstract}
 \centerline{ Brief recollections by the author how he interacted with Steven Weinberg over quantum theory.}
\end{abstract}

\title{Recollections of Steven Weinberg}

\author{James  Hartle}
\affiliation{Department of Physics, University of California, Santa Barbara, California  93106, USA} {\affiliation{ Santa Fe Institute, \\ 1399 Hyde Park Road,  Santa Fe, New Mexico  87501, USA.}  

\bibliographystyle{unsrt}
\bibliography{references}


\maketitle

\section{Introduction}
\label{intro}
The late Steven Weinberg must be regarded as one of the giants of theoretical physics for his work on a viable theory of the elementary particles and cosmology. But he was also interested in what can be called the foundations of quantum theory.   He thought that its present formulations were inadequate because they posited a rule for quantum probabilities rather than deriving the rule from a more basic formalism as in other areas of physics  and described in more detail below.

In this article I recall my interactions with Steve on the subject of quantum theory.  I am relying on my memory  for these knowing  all the risks that go along with that.  I am not writing accurate history that can be backed up by documents.  I am writing  about my recollections of situations and events that took place some time  ago.
 
   \section{Born's Rule}
   \label{brule}
   
   I had only a few opportunities to visit Texas and talk to Steve in person.  But we did discuss issues by phone and e-mail. 
   It became clear in these discussions that Steve thought that Born's rule should not  be posited but rather emerge from counting incidents, possibly more like statistical mechanics, See, e.g.  ( S. Weinberg,  {\it  The trouble with Quantum Mechanics}, {\sl New York Review} ,  Jan 19,2017).
   
   \section{Jim, is this a good time to talk?}
   \label{tmetalk} 
   After one discussion Steve said he would go off and provide a derivation of Born's rule from  the principles of quantum mechanics. 
   I told him he was never going to succeed because quantum theory is linear.  A few weeks later he sent me a { {\it non-linear} formulation of quantum mechanics with which he could derive Born's rule. I  wrote back with both praise and a list of objections. 
   
   A few days later our phone rang at home in the middle of dinner.  It was Steve.  ``Jim, is this a good time to talk?'' (about those objections) he asked.  Of course, it was always a good time to talk to Steve --- even in the middle of dinner!
   
  I  had  derived Born's rule earlier from simple assumptions in ( \it Quantum Mechanics of Individual Systems} , {\sl Am. J. Phys}. , {\bf 36}, 704-712, (1968)).  I think that Steve liked this derivation because   it's laid out straight down the party line in Section 3.7 of   his  {\sl  Lectures on Quantum Mechanics 2nd edition.(CUP, 2015).} But still he hoped for. something better.
 
 \section{Testing Quantum Theory}
 Not everything that I learned from Steve came from our conversations and e-mail exchanges.  I also got ideas from reading his books. I might mention in particular his suggestion that testing quantum mechanics would be helped if we had theories that were close to quantum mechanics but not quantum mechanics exactly. 
 See,  (Steven Weinberg, {\it Dreams of a Final Theory} , Chapter 4,  p.85} in my copy, (Pantheon Books, 1992 ).
My generalized quantum mechanics, arXiv: 2110.11628 makes this possible.  I intend to  use it to construct  such theories.

\section{Conclusion}
\label{ conclusion]}
Steve's untimely death was a significant loss for me scientifically. I had found a colleague who was both sympathetic and critical of my work.  There was every reason to believe that progress and clarity would have emerged from  common efforts.

\section{Acknowledgements} 
\vskip .1in 
\noindent{\bf Acknowledgments:}    Thanks  are due to the NSF for supporting the preparation of this essay  under grant PHY-18-8018105 and to Mary Jo Hartle for proofreading it more than once.

 \end{document}